\documentclass[11pt]{article}%
\usepackage{amsfonts,bm}
\usepackage{mathrsfs}
\usepackage{amsmath}
\usepackage{graphicx}
\usepackage{hyperref}
\usepackage{verbatim}
\usepackage{color}%
\usepackage{amsfonts}%
\usepackage{amssymb}
\providecommand{\U}[1]{\protect\rule{.1in}{.1in}}
\csname @addtoreset\endcsname{equation}{section}
\newcommand{\beq}{\begin{equation}}
\newcommand{\eeq}{\end{equation}}
\newcommand{\bea}{\begin{eqnarray}}
\newcommand{\eea}{\end{eqnarray}}
\newcommand{\ba}{\begin{array}}
\newcommand{\ea}{\end{array}}
\newcommand{\bit}{\begin{itemize}}
\newcommand{\eit}{\end{itemize}}

\newcommand{\complesso}{{\ \hbox{{\rm I}\kern-.6em\hbox{\bf C}}}}
\newcommand{\reale}{{\hbox{{\rm I}\kern-.2em\hbox{\rm R}}}}

\newcommand{\1}{ \,  \raisebox{+0.14em}{{\hbox{{\rm \scriptsize ]}} \raisebox{-0.2em}{\kern-.8em\hbox{1}}}} \, }

\begin{document}
\begin{titlepage}
\begin{flushright}
\end{flushright}
\vspace{2.6cm}
\begin{center}
\renewcommand{\thefootnote}{\fnsymbol{footnote}}
{\LARGE \bf  Exact meron Black Holes in four dimensional $SU(2)$ Einstein-Yang-Mills theory }
\vskip 20mm
{\large {Fabrizio Canfora$^{1,2}$\footnote{canfora@cecs.cl}, Francisco Correa$^{1}$\footnote{correa@cecs.cl} Alex Giacomini$^{3}$\footnote{alexgiacomini@uach.cl}} Julio Oliva$^{3,4}$\footnote{julio.oliva@docentes.uach.cl} }\\
\renewcommand{\thefootnote}{\arabic{footnote}}
\setcounter{footnote}{0}
\vskip 10mm
\textit{
$^{1}$Centro de Estudios Cient\'{\i}ficos (CECs), Valdivia, Chile.   \\
\vskip 1mm
${^2}$Universidad Andr\'es Bello, Av. Rep\'ublica 440, Santiago, Chile.  \\
\vskip 1mm
$^{3}$Instituto de Ciencias F\'{\i}sicas y Matem\'{a}ticas, Universidad Austral de Chile, Valdivia, Chile. \\
\vskip 1mm
$^{4}$Universidad de Buenos Aires FCEN-UBA, Ciudad Universitaria, Pabell\'{o}n I, 1428, Buenos Aires, Argentina.\\
}
\end{center}
\begin{center}
{\bf Abstract}
\end{center}
In this paper an intrinsically non-Abelian black hole solution for the $SU(2)$ Einstein-Yang-Mills theory in four dimensions is constructed. The gauge field of this solution has
the form of a meron whereas the metric is the one of a Reissner-Nordstr\"om black hole in which, however, the coefficient of the $1/r^2$ term is not an integration constant.
Even if the stress-energy tensor of the Yang-Mills field is spherically symmetric, the field strength of the Yang-Mills
field itself is not.
A remarkable consequence of this fact, which allows to distinguish the present solution from
essentially Abelian configurations, is the \textit{Jackiw, Rebbi, Hasenfratz, 't Hooft} mechanism
according to which excitations of bosonic fields moving in the background of a gauge field with this characteristic
behave as Fermionic degrees of freedom.  \\
\end{titlepage}


\section{Introduction}

The Yang-Mills (YM) action is one of the main ingredients of the standard
model which up to now has been phenomenologically extremely successful.
According to General Relativity, a Yang-Mills field contributes as any other
field to the curvature of the space-time. There are several physically
relevant situations (for instance, close to a neutron star or a black hole, in
the early cosmology) where the gravitational fields are extremely strong and
the effects of curvature on the propagation of matter fields, as well as the
back-reaction of the Yang-Mills fields cannot be neglected.\newline The self
gravitating Yang-Mills field is of great theoretical interest in black hole
physics. Indeed, non-Abelian gauge fields are known to violate the celebrated
\textquotedblleft no hair conjecture". This means that there exist black hole
configurations with a non-Abelian gauge field which however does not
contribute to the conserved charges \cite{Bizon}. At least in four dimensional
Einstein-YM theory such hairy black holes have been found only numerically. It
has been shown that there exist also $SU(2)$ Reissner-Nordstr\"{o}m like black
holes \cite{smoller-wasserman}.\newline Up to now in the Einstein-YM system,
in four dimensions, in spite of the great effort in constructing numerical
solutions \cite{Bizon} and in making rigorous proofs of existence of genuine
non-Abelian solutions \cite{smoller-wasserman} very few exact, intrinsically
non-Abelian solutions are known (for two detailed reviews, see refs.
\cite{volkov} and \cite{winstanley}). In particular, within the family
considered in \cite{Bizon} and \cite{smoller-wasserman} only one exact
solution of a Reissner-Nordstr\"{o}m black hole can be constructed in which
the corresponding Yang-Mills field is gauge equivalent to a potential with
only one of its three $SU(2)$ generators switched on. This means that the
mentioned solution actually belongs to an Abelian sector of the
theory.\newline On the other hand, it would be of great importance to have an
exact Yang-Mills black hole solution which is genuinely non-Abelian and which
therefore captures the most relevant characteristic features of Yang-Mills
theory, since many of the available results are numerical \cite{Bizon},
\cite{ref1}\cite{ref2}\cite{ref3}\cite{ref4}. \newline A good strategy to
construct non-Abelian black holes is to consider an ansatz for the Yang-Mills
field which is both intrinsically non-Abelian and as simple as possible: in
this paper we will consider the meron ansatz. A meron is a field configuration
which has the form $\mathbf{A}=\lambda\mathbf{\widetilde{A}}$ where
$\mathbf{\widetilde{A}}$ is pure gauge field. In an Abelian theory a multiple
of a pure gauge field is, of course, a pure gauge field as well. In a
non-Abelian theory however, the field strength has also the commutator term.
Hence, when $\lambda\neq0,1$, the meron configuration has a non-zero field
strength: $\mathbf{F}=\lambda(\lambda-1)[\mathbf{\widetilde{A}}%
,\mathbf{\widetilde{A}}]$. Therefore, the existence of merons is a genuine
non-Abelian feature.\newline Merons, firstly introduced in \cite{merons1}, are
configurations of the YM theory which in flat space-time attracted a lot of
attention. They interpolate between different topological sectors and, in
particular, it can be shown that instantons can be thought of as composed by a
pair of a meron and an anti-meron \cite{meronsconfinement1}
\cite{meronsconfinement2} \cite{meronsconfinement3} \cite{meronsconfinement4}.
Furthermore, at least on flat spaces, merons are quite relevant configurations
as far as confinement is concerned \cite{meronsconfinement1}
\cite{meronsconfinement2} (for a recent discussion see also
\cite{meronsconfinement3}). It is also worth noting that the existence of
merons is closely related to the presence of Gribov copies discovered in the
seminal paper \cite{Gri78}, in fact, it has been shown that one can interpret
a meron as a tunneling between a two Gribov vacua \footnote{See the reviews
\cite{meronsconfinement4} and \cite{Sobreiro-Sorella}}. Due to the fact that
the pattern of appearance of Gribov copies on curved space-time may be quite
different form the flat case as it has been shown in \cite{CGO} \cite{ACGO}
\cite{CGO2}, it is natural to analyze how the curvature of space-time affects
the presence of merons.

Merons are also important for another reason..A very deep feature of
non-Abelian gauge theories first noticed in \cite{gaugecopy1}
\cite{gaugecopy2} is that, unlike what happens in Abelian theories, the
non-Abelian field strength does not uniquely determine the non-Abelian gauge
potential modulo gauge transformations and it is possible to construct many
examples of \textit{gauge potentials which are not gauge equivalent but have
the same field strength }(of course, they are distinguished by higher order
invariants). It has been shown in \cite{gaugecopy2} that, using merons, it is
easy to construct examples of non-equivalent gauge potential with the same
curvature.\newline In this paper we will construct an analytic black hole
corresponding to the energy-momentum tensor of a meron in the case where the
constant $\lambda$ takes the value $1/2$, which turns out to be reminiscent of
the original paper of de Alfaro, Fubini and Furlan \cite{merons1}. The metric
of the solution will be the one of a magnetically charged
Reissner-Nordstr\"{o}m black hole. Nevertheless, we will show that it is
impossible to transform our meron black hole into the known analytic solution
in \cite{Bizon} and \cite{smoller-wasserman}\ (which belongs to an Abelian
sector) by any globally defined $SU(2)$ gauge transformation.

Furthermore, it is possible to disclose the genuine non-Abelian nature of the
present black hole solution with a non-trivial physical effect. As it will be
shown in the next sections, the Yang-Mills stress tensor is spherically
symmetric but the field strength itself is not (unless one compensates a
spatial rotation with an internal $SU(2)$ rotation). This fact is the physical
origin of the \textit{Jackiv-Rebbi-Hasenfratz-'t\ Hooft\ effect} \cite{JR}
\cite{tH} according to which excitations of Bosonic fields charged under
$SU(2)$ around the meron black hole solution are Fermionic despite to the fact
that all the fundamental fields involved in the model are Bosonic. This
phenomenon is not restricted to Yang-Mills theory: the earliest and most
famous example is probably the (Bosonic) Skyrme field \cite{skyrme} (for a
detailed review see \cite{manton}). Indeed, the excitations around the Skyrme
soliton behaves as Fermions.\newline This gives the possibility of physically
distinguishing our solutions from the analytic solution of the abelian sector.

The structure of the paper will be the following. In the second section, a
short review of the merons will be presented. The third section is devoted to
the discussion of the hedgehog ansatz. In section four the meron black hole
will be analyzed. In the fifth section it will be shown that the present meron
black hole is not continuously connected to any Abelian sector, the
non-Abelian charges will be analyzed as well. In the sixth section, we will
show that the \textit{Jackiv-Rebbi-Hasenfratz-'t\ Hooft} mechanism is a nice
observable effect able to distinguish the present black hole solution from an
Abelian solution. In the final section some conclusions will be drawn.


\section{A Short Review on Merons}


One of the most important features of Yang-Mills theory is the presence of
topologically non-trivial configurations such as instantons, merons, monopoles
and so on (see, for instance, \cite{duality}). In the present section we will
focus on the computations of the energy-momentum tensor of the merons as well
as of the corresponding Yang-Mills equations. Let us consider the following
action $S_{YM}$ for the Yang-Mills system for the gauge group $SU(2)$,%
\begin{equation}
S_{YM}=\frac{1}{2e^{2}}\int\sqrt{-g}\, d^{4}x\, Tr\left(  F_{\mu\nu}F^{\mu\nu
}\right)  \, ,\label{symh}%
\end{equation}
where $g$ is the determinant of the metric tensor and
\begin{align*}
A_{\mu}  & =iA_{\mu}^{i}\sigma_{i}\ ,\\
\sigma_{i}\sigma_{j}  & =\delta_{ij}\mathbf{1}+i\varepsilon_{ijk}\sigma_{k} \,
,\\
F_{\mu\nu}  & =\partial_{\mu}A_{\nu}-\partial_{\nu}A_{\mu}+\left[  A_{\mu
},A_{\nu}\right]  \, .
\end{align*}
Here $\sigma_{i}$ are the Pauli matrices that we have choose as the Hermitian
generators of $su(2)$ ($[\sigma_{i},\sigma_{j}]=2i\varepsilon_{ijk}\sigma_{k}%
$) and $\mathbf{1}$ is the $2\times2$ identity matrix. $e$ is the coupling
constant and the Latin letters ($i$, $j$, $k$) correspond to the gauge group
indices. $\varepsilon_{ijk}$\ is the Levi-Civita symbol that fulfills the
identity $\varepsilon_{ijk}\varepsilon_{mnk}=\delta_{im}\delta_{jn}%
-\delta_{in}\delta_{jm}$.

As mentioned in the introduction, a meron is a configuration of the following
form,%
\begin{align}
A_{\mu}  & =\lambda U^{-1}\partial_{\mu}U, \qquad\lambda\neq0,1\,
,\label{meronsansatz1}\\
U  & =U(x^{\mu})\in SU(2)\, .\nonumber
\end{align}
Thus, a meron is proportional to a pure gauge term without being, of course, a
pure gauge configuration. It is worth emphasizing that the existence of merons
is an \textit{intrinsically non-Abelian feature} since, obviously, in an
Abelian gauge theory a gauge field which is proportional to a pure gauge is
itself a pure gauge. Thus, merons only exist in non-Abelian sectors of gauge
theories\footnote{This fact has the following practical advantage: when one
searches for exact solutions of the Einstein-Yang-Mills system, there is
always the risk that, by simplifying too much the gauge potential, at the end
one reduces $A_{\mu} $ to an Abelian gauge field (namely, a configuration in
which the commutator in the field strength vanishes). This point will be
analyzed in more details in the next sections.}.

The most famous meron configuration on flat space-time have been constructed
by de Alfaro, Fubini and Furlan \cite{merons1} and it has $\lambda=\frac{1}%
{2}$. In principle $\lambda$ could take any value different from zero and one.
However, using a purely topological argument, we will show why $\lambda=1/2$
is indeed a special value, even in curved space-time. Soon afterwords its
discovery, it was recognized that merons are very important to explain, at
least at a qualitative level, confinement \cite{meronsconfinement1}
\cite{meronsconfinement2}. The close relations between merons and confinement
has been recently confirmed in \cite{meronsconfinement3} (for a review of the
original arguments see \cite{meronsconfinement4}).

The field strength $F_{\mu\nu}$ of the meron in Eq. (\ref{meronsansatz1}) is
proportional to the commutator,%
\begin{equation}
F_{\mu\nu}=\lambda\left(  \lambda-1\right)  \left[  U^{-1}\partial_{\mu
}U,U^{-1}\partial_{\nu}U\right]  \, .\label{curvamerons}%
\end{equation}
In the following we will use the following standard parametrization of the
$SU(2)$-valued functions $U(x^{\mu})$:
\begin{align}
U(x^{\mu})  & =Y^{0}\mathbf{1}+i \, Y^{i}\sigma_{i}, \qquad U^{-1}(x^{\mu
})=Y^{0}\mathbf{1}-iY^{i}\sigma_{i}\, ,\label{standard1}\\
Y^{0}  & =Y^{0}(x^{\mu}), \qquad Y^{i}=Y^{i}(x^{\mu})\ ,\label{standard2}\\
\left(  Y^{0}\right)  ^{2}+Y^{i}Y_{i}  & =1\ ,\label{standard3}%
\end{align}
where, the sum over repeated indices is understood also in the case of the
group indices (in which case the indices are raised and lowered with the flat
metric $\delta_{ij}$). Therefore, the meron gauge field in Eq.
(\ref{meronsansatz1}) can be written as follows,
\begin{align}
A_{\mu}  & =i \,\lambda P_{\mu}^{k}\sigma_{k}\, ,\label{standard4}\\
P_{\mu}^{k}  & =\varepsilon_{ijk}Y_{i}\partial_{\mu}Y_{j}+Y^{0}\partial_{\mu
}Y^{k}-Y^{k}\partial_{\mu}Y^{0}\, .\label{standard5}%
\end{align}
In order to determine the energy-momentum tensor of the meron field it is
useful to compute the following quadratic combination,
\begin{align}
\delta_{mn}P_{\mu}^{m}P_{\alpha}^{n}  & =G_{ij}\partial_{\mu}Y^{i}%
\partial_{\alpha}Y^{j}\, ,\label{cuadra1}\\
G_{ij}  & =G_{ij}\left(  \overrightarrow{Y}\right)  =\left(  \delta_{ij}%
+\frac{Y_{i}Y_{j}}{1-Y^{k}Y_{k}}\right)  \, ,\label{cuadra2}%
\end{align}
where $G_{ij}$ is the metric corresponding to the group manifold which, in the
present case is $S^{3}$.
It is worth to note here that if one considers a configuration in which
$Y^{0}$ vanishes, then the internal metric $G_{ij}\left(  \overrightarrow
{Y}\right) $ reduces to the $\delta_{ij}$,
\begin{align}
Y^{0}  & =0 \quad\Rightarrow\quad Y^{k}Y_{k}=1 \quad\Rightarrow\label{part1}\\
G_{ij}\left(  \overrightarrow{Y}\right)   & =\delta_{ij}\, .\label{part2}%
\end{align}

The energy-momentum tensor for the Yang-Mills field reads
\begin{equation}
T_{\mu\nu}=\frac{1}{e^{2}}Tr\left(  -F_{\mu\alpha}F_{\nu\beta}g^{\alpha\beta
}+\frac{g_{\mu\nu}}{4}F_{\lambda\sigma}F^{\lambda\sigma}\right)  \,
,\label{tmunu1}%
\end{equation}
and using Eqs. (\ref{cuadra1}) and (\ref{cuadra2}) the energy-momentum tensor
for the merons reduces to
\begin{align}
T_{\mu\nu}  & =\xi\left[  \left(  g^{\alpha\beta}G_{ij}\partial_{\alpha}%
Y^{i}\partial_{\beta}Y^{j}\right)  G_{mn}\partial_{\mu}Y^{m}\partial_{\nu
}Y^{n}-g^{\alpha\beta}\left(  G_{ij}G_{mn}\partial_{\mu}Y^{i}\partial_{\beta
}Y^{j}\partial_{\nu}Y^{m}\partial_{\alpha}Y^{n}\right)  \right.  +\nonumber\\
& \left.  -\frac{g_{\mu\nu}}{4}\left(  \left(  g^{\alpha\beta}G_{ij}%
\partial_{\alpha}Y^{i}\partial_{\beta}Y^{j}\right)  ^{2}-g^{\alpha\beta
}g^{\lambda\sigma}\left(  G_{ij}G_{mn}\partial_{\lambda}Y^{i}\partial_{\beta
}Y^{j}\partial_{\sigma}Y^{m}\partial_{\alpha}Y^{n}\right)  \right)  \right]
\, ,\label{tmunu2}%
\end{align}
where
\[
\xi=\frac{8\left(  \lambda(\lambda-1\right)  )^{2}}{e^{2}}\, .
\]
Finally, the Yang-Mills equations for the meron field read,%
\begin{equation}
\varepsilon_{lmn}\nabla^{\nu}\left(  P_{\mu}^{m}P_{\nu}^{n}\right)
-2\lambda\left(  \varepsilon_{kjm}\varepsilon_{lik}\right)  P^{i\nu}P_{\mu
}^{j}P_{\nu}^{m}=0\, .\label{ymequations}%
\end{equation}

\section{The Hedgehog ansatz}

In the following we will consider the spherically symmetric hedgehog ansatz
for the meron field in terms of a group valued function $U$. The notion of
spherical symmetry in which one gets \textit{spherical symmetry only up to an
internal} $SU(2)$\textit{\ rotation} is the one introduced\footnote{This
definition is locally but not globally equivalent to the one which is commonly
adopted in the analysis of colored black holes.} in \cite{teitelboim} (see
also \cite{witten}) and, as it will be explained in the section \ref{52}, it
is responsible for the appearance of the Jackiw-Rebbi-Hasenfratz-'t Hooft
effect \cite{JR} \cite{tH}. In terms of the group element $U$ it reads
\begin{align}
U  & =\mathbf{1}\cos f(r)+i\,\widehat{x}^{i}\sigma_{i}\sin f(r), \quad
U^{-1}=\mathbf{1}\cos f(r)-i\,\widehat{x}^{i}\sigma_{i}\sin f(r)\,
,\label{hedgehog1}\\
\delta_{ij}\widehat{x}^{i}\widehat{x}^{j}  & =1\, ,\nonumber
\end{align}
where $\widehat{x}^{j}$ is the unit radial vector (normalized with respect to
the internal metric $\delta_{ij}$). The hedgehog ansatz corresponds to the
following choice,
\begin{align}
Y^{0}  & =\cos f(r), \quad Y^{i}=\widehat{x}^{i}\sin f(r)\, ,\nonumber\\
\widehat{x}^{1}  & =\sin\theta\cos\phi, \quad\widehat{x}^{2}=\sin\theta
\sin\phi, \quad\widehat{x}^{3}=\cos\theta\ .\label{single}%
\end{align}
Thus, the meron gauge field in Eq. (\ref{standard4}) in this case reads as
follows,%
\begin{align}
A_{\mu}  & =i\lambda P_{\mu}^{k}\sigma_{k}\, ,\label{hedgehog1.5}\\
P_{\mu}^{k}  & =\sin^{2}f\varepsilon_{ijk}\widehat{x}^{i}\partial_{\mu
}\widehat{x}^{j}+\widehat{x}^{k}\partial_{\mu}f+\frac{\sin\left(  2f\right)
}{2}\partial_{\mu}\widehat{x}^{k}\, .\label{hedgehog2}%
\end{align}
As it will be discussed in the next section, one can obtain an exact solution
of the Einstein-Yang-Mills system in the case in which
\begin{equation}
f(r)=\frac{\pi}{2}\ \Rightarrow P_{\mu}^{k}=\varepsilon_{ijk}\widehat{x}%
^{i}\partial_{\mu}\widehat{x}^{j}\, .\label{RN00}%
\end{equation}
In this case the field strength $F_{\mu\nu}=iF_{\mu\nu}^{k}\sigma_{k}$ of the
non-Abelian field is purely magnetic and reads
\begin{align}
F_{\mu\nu}^{i}  & =Y^{i}\Pi_{\mu\nu}\, ,\label{RN01}\\
\Pi_{\mu\nu}  & =2\lambda(\lambda-1)\left(  \varepsilon_{mnq}Y^{m}%
\partial_{\left[  \mu\right.  }Y^{q}\partial_{\left.  \nu\right]  }%
Y^{n}\right) \, ,\label{RN02}%
\end{align}
so we define the two form $\Pi$,
\[
\Pi:=\frac{1}{2}\Pi_{\mu\nu}dx^{\mu}\wedge dx^{\nu}=\lambda(\lambda
-1)\varepsilon_{mnq}Y^{m}dY^{q}\wedge dY^{n}\ ,
\]
with $\delta_{ij}Y^{i}Y^{j}=1$. The functions $Y^{i}$ are define since Eq.
(\ref{RN00}) which implies,
\begin{align}
Y^{1}  & =\widehat{x}^{1}=\sin\theta\cos\phi\, ,\ \nonumber\\
Y^{2}  & =\widehat{x}^{2}=\sin\theta\sin\phi\, ,\nonumber\\
Y^{3}  & =\widehat{x}^{3}=\cos\theta\, ,\label{spacc}%
\end{align}
where $\theta$\ and $\phi$\ are the coordinates on the two sphere
corresponding to the metric in Eq. (\ref{metrcoulomb}).

\subsection{The geometrical meaning of $\Pi_{\mu\nu}$}

It is worth emphasizing that $\Pi_{\mu\nu}$ has the same form as an effective
Abelian magnetic field strength. It is easy to see that the energy-momentum
tensor corresponding to the non-Abelian field strength in Eq. (\ref{RN01})
coincides with twice the (Maxwell) energy-momentum tensor of $\Pi_{\mu\nu}$.
This is due to the fact that the trace over the group indices in the
energy-momentum tensor eliminates the explicit factor $Y^{i}$ which multiplies
$\Pi_{\mu\nu}$ in Eq. (\ref{RN01}) thanks to $Y_{i}Y_{i}=1$. Thus,
\[
T_{\mu\nu}=\frac{1}{e^{2}}Tr(-g^{\alpha\beta}F_{\mu\alpha}F_{\nu\beta}%
+\frac{g_{\mu\nu}}{4}F_{\alpha\beta}F^{\alpha\beta})\ ,
\]
reduces to
\[
T_{\mu\nu}=\frac{2}{e^{2}}(\Pi_{\mu\alpha}\Pi_{\nu\beta}-\frac{g_{\mu\nu}}%
{4}\Pi_{\alpha\beta}\Pi^{\alpha\beta})\ ,
\]

For any triple of functions $Y^{i}$ satisfying the relation in $Y_{i}Y_{i}=1$
the expression in Eq. (\ref{RN02}) represents the pull-back of the area form
on $S^{2}$ and its integral represents the $\pi_{3}(S^{2})$. This implies that
the two-form $\Pi$ is closed,%
\begin{align*}
d\Pi & =\lambda(\lambda-1)d\left(  \varepsilon_{mnq}Y^{m}dY^{q}\wedge
dY^{n}\right)  =0\ \Rightarrow\\
\Pi & =dA\ \ \ \ locally\ ,
\end{align*}
and, therefore, $\Pi_{\mu\nu}$ satisfies the first set of Maxwell equations.
Furthermore, as it is well known, the field strength $F_{D}$ of the Dirac
monopole reads%
\begin{equation}
F_{D}=\frac{g}{4\pi}\sin\theta d\theta\wedge d\phi\ ,\label{dirac1}%
\end{equation}
where $g$ is the magnetic charge and the field strength $F_{D}$ is
proportional to the volume form of $S^{2}$. Thus, it turns out that the
effective Abelian field defined in Eq. (\ref{RN02}) is proportional to $F_{D}
$,%
\begin{equation}
\frac{\Pi}{2\lambda(\lambda-1)}=-\frac{4\pi}{g}F_{D}=-\sin{\theta}d{\theta
}\wedge d{\phi}\ ,\label{dirac4}%
\end{equation}
and so the effective Abelian field strength $\Pi_{\mu\nu}$ defined in Eq.
(\ref{RN02}) automatically satisfies also the second set of Maxwell equations
with a $\delta$-like source.

\section{The Black Hole solution}

The Einstein equations derived from the action,
\begin{equation}
S[g_{\mu\nu},A_{\mu}]=\int d^{4}x\sqrt{-g}\left(  \frac{R-2\Lambda}{\kappa
}+\frac{1}{2e^{2}}Tr\left(  F_{\mu\nu}F^{\mu\nu}\right)  \right)  \ ,
\end{equation}
read
\begin{equation}
G_{\mu\nu}+\Lambda g_{\mu\nu}=\frac{\kappa}{e^{2}}T_{\mu\nu}\ ,
\end{equation}
($\kappa$ and $e$ being the Newton and Yang-Mills coupling constants
respectively). The energy momentum tensor $T_{\mu\nu}$ is given in Eq.
(\ref{tmunu1}). Let us consider a four-dimensional metric of the form
\begin{align}
ds^{2}  & =-\exp\left(  2a(r)\right)  dt^{2}+\exp\left(  2b(r)\right)
dr^{2}+r^{2}\left(  d\theta^{2}+\sin^{2}\theta d\phi^{2}\right)
\ ,\label{metrcoulomb}\\
0  & \leq r<\infty\ ,\ \ \ 0\leq t<\infty\ \ .\nonumber
\end{align}
Since, as it has been already explained, the energy-momentum tensor
corresponding to the above meron field strength in Eqs. (\ref{RN01}),
(\ref{RN02})\ coincides with the energy-momentum tensor of a Dirac monopole,
the coupled Einstein-Yang-Mills system of equations (both with and without
cosmological constant) is solved, for any value of the meron parameter
$\lambda$, by the magnetic Reissner-Nordstrom black hole metric%
\begin{equation}
ds^{2}=-\left(  1-\frac{\kappa M}{8\pi r}+\frac{4\kappa\lambda^{2}%
(\lambda-1)^{2}}{e^{2}r^{2}}-\frac{\Lambda r^{2}}{3}\right)  dt^{2}%
+\frac{dr^{2}}{1-\frac{\kappa M}{8\pi r}+\frac{4\kappa\lambda^{2}%
(\lambda-1)^{2}}{e^{2}r^{2}}-\frac{\Lambda r^{2}}{3}}+r^{2}\left(  d\theta
^{2}+\sin^{2}\theta d\phi^{2}\right)  \ .\label{bhsolution}%
\end{equation}
where $M$ is the ADM mass. However, the Yang-Mills equations have not been
solved yet: it will be now shown that the Yang-Mills equations corresponding
to the present meron ansatz fix the value of $\lambda$ as in the original de
Alfaro-Fubini-Furlan paper, namely $\lambda=1/2$. Hence, unlike the Abelian
case, $\lambda$ is not an integration constant but is fixed to be $1/2$.

The Yang-Mills equations read:%
\begin{align*}
Y\!\!M_{\mu}=\nabla^{\nu}F_{\mu\nu}+\left[  A^{\nu},F_{\mu\nu}\right]   &
=0\ ,\\
\left[  A^{\nu},F_{\mu\nu}\right]  ^{i}  & =-2i\lambda\Pi_{\mu\nu}\nabla^{\nu
}Y^{i}\ ,
\end{align*}
($\nabla^{\nu}$ being the Levi-Civita covariant derivative corresponding to
the metric in Eq. (\ref{metrcoulomb})) so that they reduce to
\begin{equation}
\Pi_{\mu\nu}\nabla^{\nu}Y^{i}-2\lambda\nabla^{\nu}\left(  \Pi_{\mu\nu}%
Y^{i}\right)  =0\ .\label{multipleYangMills}%
\end{equation}
Furthermore, $\Pi_{\mu\nu}$ is proportional to the field strength of the Dirac
monopole
and so it satisfies the Maxwell equations (outside the $\delta-$source)
\[
\nabla^{\nu}\Pi_{\mu\nu}=0\ ,
\]
therefore Eq. (\ref{multipleYangMills}) can be written as
\begin{equation}
\left(  1-2\lambda\right)  \Pi_{\mu\nu}\nabla^{\nu}Y^{i}%
=0\ ,\label{finalyangmills}%
\end{equation}
whose non-trivial components are
\begin{align}
Y\!\!M_{\theta}^{1}=r^{-2}\sin\phi\left(  \lambda-\frac{1}{2}\right)  ,\quad
Y\!\!M_{\phi}^{1}  & =r^{-2}\cos\phi\sin\theta\cos\theta\left(  \lambda
-\frac{1}{2}\right)  \,,\label{ym1}\\
Y\!\!M_{\theta}^{2}=r^{-2}\cos\phi\left(  \lambda-\frac{1}{2}\right)  ,\quad
Y\!\!M_{\phi}^{2}  & =r^{-2}\cos\phi\sin\theta\cos\theta\left(  \lambda
-\frac{1}{2}\right)  \,,\label{ym2}\\
Y\!\!M_{\phi}^{3}  & =r^{-2}\sin^{2}\theta\left(  \lambda-\frac{1}{2}\right)
\,.\label{ym3}%
\end{align}
Therefore, all the above equations are simultaneously satisfied if and only
if
\begin{equation}
\lambda=\frac{1}{2}\,.\label{RN05}%
\end{equation}
To the best of authors knowledge, the above argument provides an additional
explanation of why the value $\lambda=1/2$ is special in the present
Lorentzian meron which is reminiscent of the more well-known Euclidean ones.

It is worth emphasizing that, even on flat spaces, merons present
singularities (see, for instance, \cite{meronsconfinement1}
\cite{meronsconfinement2} \cite{meronsconfinement4}) and so they play an
important although indirect role as building blocks of the instantons but they
cannot be observed directly due to their singularities (which, in the present
case, are manifest in Eq. (\ref{singomeron})). However, one of the most
interesting results of the present analysis is that the meron singularity is
hidden behind the black hole horizon and, consequently, in a gravitational
context, merons could be observed directly in principle.

The causal structure of the black hole solution corresponds to the one of the
Reissner-Nordstr\"om-(A)dS space-time. However, an essential difference
between the present and the Abelian Reissner-Nordstr\"om solutions is that in
this case the coefficient of the $1/r^{2}$ term in the lapse function is not
an integration constant. Its value is fixed to $1/4$ which is the square of
the non-Abelian magnetic charge of the configuration.

For $\Lambda=-\frac{3}{l^{2}}<0$, there is an event horizon provided $M\geq
M_{c}$, where $M_{c}$ is a critical mass. $M_{c}$ is related with the minimum
radius of the horizon $r_{c}$ as follows,
\[
r_{c}^{2}=\frac{l^{2}}{6}\left[  \sqrt{1+\frac{3\kappa}{e^{2}l^{2}}}-1\right]
,\quad M_{c}=\frac{16\pi r_{c}}{k}\left[  1+\frac{2r_{c}^{2}}{l^{2}}\right]
\ .
\]
When the lower bound of the mass is achieved, the event and the Cauchy
horizons coincide and the black hole is extremal. In the asymptotically flat
case ($\Lambda=0$) the causal structure is similar to the asymptotically AdS
case with
\[
r_{c}^{flat}=\frac{\sqrt{\kappa}}{2e},\quad M_{c}^{flat}=\frac{4\pi}%
{e^{2}r_{c}^{flat}}\ .
\]
Finally, in the asymptotically de Sitter case (with $\Lambda=\frac{3}{l^{2}%
}>0$), if we define $r_{min}$ and $r_{max}$ respectively by,
\[
r_{\min}^{2}=\frac{l^{2}}{6}\left[  1-\sqrt{1-\frac{3\kappa}{e^{2}l^{2}}%
}\right]  ,\quad r_{\max}^{2}=\frac{l^{2}}{6}\left[  1+\sqrt{1-\frac{3\kappa
}{e^{2}l^{2}}}\right]  \,
\]
the minimum and maximum masses read
\[
M_{\min}=\frac{16\pi r_{\min}}{k}\left[  1-\frac{2r_{\min}^{2}}{l^{2}}\right]
,\quad M_{\max}=\frac{16\pi r_{\max}}{k}\left[  1-\frac{2r_{\max}^{2}}{l^{2}%
}\right]  \,.
\]
When $M<M_{min}$ the space-time represents a naked singularity. For
$M=M_{min}$ the space-time represents an extremal black hole surrounded by a
cosmological horizon, giving rise to what is known as a \textquotedblleft
lukewarm" black hole. When $M_{min}<M<M_{max}$ there is a Cauchy and an event
horizon, both surrounded by the cosmological horizon. Finally for $M=M_{max}$
the event and the cosmological horizons coincide. For masses above this value,
the space-time represents again a naked singularity.

\section{The present meron-black hole as a genuine non-Abelian configuration}

We have seen that both the metric of the meron black hole and the
corresponding energy-momentum tensor look like a magnetically charged Reissner
Nordstrom black hole. This fact, at a first glance, may give the impression
that this solution is gauge equivalent to an Abelian solution.

In particular, within the family of configurations analyzed in
\cite{smoller-wasserman} to prove non-trivial existence theorems (as well as
in \cite{Bizon}\ to construct numerically hairy colored black hole), there is
a configuration which allows to construct an analytic black hole solution but
it belongs to an Abelian sector of Yang-Mills theory. The corresponding
magnetic non-Abelian field strength reads
\begin{equation}
\mathbf{F}_{\mu\nu}=i\Pi_{\mu\nu}\sigma_{3}\label{smollwass1}%
\end{equation}
where $\Pi_{\mu\nu}$ is the one defined in Eq. (\ref{dirac1}) and $\sigma_{3}$
is a fixed generator of the algebra of $SU(2)$. Thus, to obtain such a field
strength it is enough to consider a gauge potential with only one generator
(namely, $\sigma_{3}$) turned on in such a way that the commutators both in
the Yang-Mills and in the Einstein equations vanish and the solution reduces
globally to an Abelian black hole.

On the other hand, as far as the solution constructed in the present paper is
concerned, the gauge field is a meron: an intrinsically non-Abelian object.
This observation by itself strongly suggests that present solution cannot be
gauge transformed to an Abelian sector. We will now present two rigorous
arguments which proves that there is no continuous gauge transformation
connecting our solution with an Abelian sector.

The first argument is the following: let us compare the field strength in Eq.
(\ref{smollwass1}) with the field strength of the present solution (see Eqs.
(\ref{RN01}) and (\ref{RN02})) which reads
\begin{align}
\mathbf{F}_{\mu\nu}  & =iF_{\mu\nu}^{i}\sigma_{i}=i\Pi_{\mu\nu}Y^{i}\sigma
_{i}\equiv i\Pi_{\mu\nu}\sigma_{r}\ ,\label{meronmeron}\\
i\Pi_{\mu\nu}Y^{i}\sigma_{i}  & \equiv i\Pi_{\mu\nu}\sigma_{r}%
\ ,\label{meronmeron2}%
\end{align}
where we have introduced the \textit{radial} Pauli matrix $\sigma_{r}\equiv
Y^{i}\sigma_{i}$ and the $Y^{i}$ are defined in Eq. (\ref{spacc}). At a first
glance, the two field strengths in Eqs. (\ref{smollwass1}) and
(\ref{meronmeron}) look similar since they are both proportional to $\Pi
_{\mu\nu}$ (which is the field strength of a Dirac monopole). However, in the
first case in Eq. (\ref{smollwass1}) $\Pi_{\mu\nu}$ multiplies a constant
generator of the algebra of $SU(2)$ while in the meron case $\Pi_{\mu\nu}$
multiplies the \textit{radial} Pauli matrix $\sigma_{r}$ which is a
non-trivial and non-constant combination of the generators of $SU(2)$.
Therefore, if the solution constructed in the present paper would be
equivalent to an Abelian configuration then one should be able to find a
smooth gauge transformation $U(x)\in SU(2)$ such that%
\begin{align}
U^{-1}\left(  \Pi_{\mu\nu}\sigma_{r}\right)  U  & =\Pi_{\mu\nu}\sigma
_{3}\ \Leftrightarrow\nonumber\\
U^{-1}\sigma_{r}U  & =\sigma_{3}\ ,\label{abeliantrans}%
\end{align}
where we have used the fact that in the expression of $\Pi_{\mu\nu}$ in Eq.
(\ref{RN02}) all the internal indices are contracted so that the gauge
transformation only acts on $\sigma_{r}$. We will now show that no such $U$
can exist.

\begin{figure}[h]
\centering
\includegraphics[scale=1]
{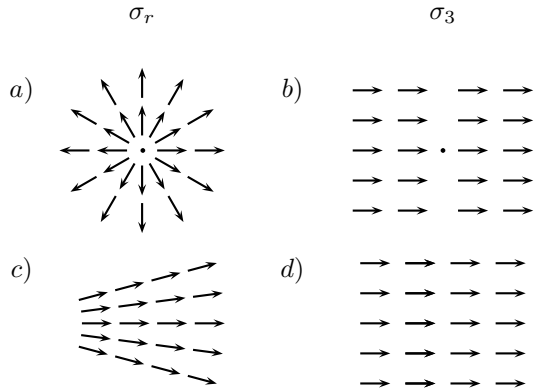}\caption{Orientation of the field strength in the isospin space.
Fig. $a)$ and $b)$ represent the neighbourhoods of the origin of the present
meron solution and an effective Abelian solution, respectively. It is not
possible to transform Fig. $a)$ into Fig. $b)$ by a local \emph{continous}
rotation. On the other hand, in a neighbourhood which does not include the
origin (Fig. $c)$ and Fig. $d)$) it is possible to transform one field
strength into the other by a local rotation. }%
\label{fig1}%
\end{figure}

The \textit{radial} Pauli matrix in Eq. (\ref{meronmeron2}) points outwards in
the radial direction of the inner space at every point of the physical space.
The $\sigma_{3}$ generator (corresponding to the field strength in Eq.
(\ref{smollwass1})) at every point of the physical space points in the same
direction of the inner space (see Fig. \ref{fig1}). Hence, if one considers a
small enough neighborhood of the origin in the physical space, one can see
that any gauge transformation $U(x,y,z)$ satisfying Eq. (\ref{abeliantrans})
is necessarily discontinuous. Indeed (introducing a Cartesian coordinates
system around the origin) one can see that, for instance, the radial Pauli
matrix in Eq. (\ref{meronmeron2}) behaves as follows,
\begin{equation}
\forall\varepsilon>0:\ \ \ \sigma_{r}(0,0,\varepsilon)=\sigma_{3}%
\mathbf{\ ,\ \ }\sigma_{r}(0,0,-\varepsilon)=\mathbf{-}\sigma_{3}%
\ ,\label{singomeron}%
\end{equation}
where $(0,0,\varepsilon)$\ and $(0,0,-\varepsilon)$\ are two points
(symmetrically placed with respect to the origin) along the $z$ axis.
Therefore, due to Eq. (\ref{abeliantrans}), one should require the following
condition on $U(x,y,z)$ (see Fig. \ref{fig1}):
\begin{align}
\forall\varepsilon & >0:\nonumber\\
\left(  U(0,0,\varepsilon)\right)  ^{-1}\sigma_{3}\left(  U(0,0,\varepsilon
)\right)   & =\sigma_{3}\mathbf{\ \ ,}\label{disc1}\\
\left(  U(0,0,-\varepsilon)\right)  ^{-1}\sigma_{3}\left(  U(0,0,-\varepsilon
)\right)   & =-\sigma_{3}\mathbf{\ \ ,}\label{disc2}%
\end{align}
so that $U(x,y,z)$ cannot be continuous since the quantity $\left(
U(0,0,\varepsilon)\right)  ^{-1}\sigma_{3}\left(  U(0,0,\varepsilon)\right)  $
is not continuous with respect to $\varepsilon$ around $\varepsilon=0$. This
obviously implies that the field strengths in Eqs. (\ref{smollwass1}) and
(\ref{meronmeron}) \textit{are not continuously connected}.

Since in the present case the origin of the coordinates is a singularity, it
is natural to wonder whether if, when removing a small spherical region around
the singularity, the gauge transformation leading from one configuration to
the other remains singular. Indeed, for topological reasons (see for instance
\cite{teitelboim}) the transformation must be singular at least on one point
of each sphere of fixed radius $r$. A more direct way to see this is the
following. Any gauge transformation which transforms the field strength in Eq.
(\ref{meronmeron}) into the one in Eq. (\ref{smollwass1}) and at the same time
the meron gauge potential in Eqs. (\ref{hedgehog1.5}), (\ref{hedgehog2}) and
(\ref{RN00}) into an Abelian gauge potential with only one generator turned is
actually singular along the whole $z-$axis and not only at the origin. The
reason is that the meron gauge potential is only singular at $r=0$ while the
gauge potential of the Dirac monopole corresponding to the field strength in
Eq. (\ref{smollwass1}) is singular along the whole $z-$axis. Therefore, any
gauge transformation of the above type must be singular along the $z-$axis as
well. 
A very famous example in General Relativity of two configurations which are connected by a gauge transformation which is not globally well defined are the $2+1$ dimensional AdS space-time and the BTZ  black hole \cite{BTZ}. This example clearly shows that improper gauge transformations have in general non trivial physical effects since $2+1$ AdS is the vacuum of the theory while the BTZ black hole has temperature and entropy and therefore they cannot be gauge equivalent.\\
A second easier way to prove that the present Einstein-Yang-Mills
configuration is genuinely non-Abelian is the following. Once one fixes
$f=\pi/2$ in Eqs.(\ref{hedgehog1.5}) and (\ref{hedgehog2}), the field strength
of the meron configuration has the form in Eqs. (\ref{RN01}) and
(\ref{meronmeron}) \textit{for any value of} $\lambda$. However, if it would
exist a regular gauge transformation transforming the present meron gauge
potential in Eqs. (\ref{hedgehog1.5}), (\ref{hedgehog2}) and (\ref{RN00}) into
an Abelian gauge potential with only one generator turned on, then one could
solve trivially the Yang-Mills equations for any value of $\lambda$ (since
they would reduce to the Maxwell equations) and then one could go back to the
meron form in Eqs. (\ref{hedgehog1.5}), (\ref{hedgehog2}) and (\ref{RN00}).
However, in the previous section we showed that the Yang-Mills equations are
satisfied if and only if $\lambda=1/2$.

\subsection{The non-Abelian charges}

The fact that the meron field and the Abelian configuration produce the same
stress tensor and therefore the same metric would lead to think that physics
is not able to distinguish between the Abelian and meron black hole
configuration. However, globally the Abelian and the non-Abelian black hole
configurations are different as it is apparent in the computations of
non-local quantities like Wilson loops (in particular, the radial Pauli matrix
at a point does not commute in general with the radial Pauli matrix at another
point). In the next subsection we will discuss a more direct physical effect
which is able to reveal the non-Abelian nature of the present solution
distinguishing it from an Abelian black hole configuration. Here we will
discuss the non-Abelian charges of the configuration.

It is worth emphasizing here that the non-Abelian charges are gauge invariant
only under \textit{proper gauge transformations} (namely, everywhere smooth
gauge transformations which approach the center of the gauge group at spatial
infinity). Indeed, in the non-Abelian case, if the gauge transformation does
not approach the center of the gauge group at infinity, the charges defined as
surface integrals of the non-Abelian fluxes at infinity may change (for a
detailed review on the concept of charges in Yang-Mills theory see
\cite{qcdcharges}). Therefore, in order for the concept of non-Abelian hair to
be well defined, the only allowed gauge transformations have to be proper
gauge transformations. Of course, any transformation which maps the field
strength (\ref{RN01}) into an Abelian one is improper for two reasons: it is
singular at the origin and it cannot approach the center of the gauge group at infinity.

The classic definition of non-Abelian charge is in \cite{abbottdeser} (see
also \cite{Jackiwtopo} \cite{mann}). The first step is to find a
$SU(2)$-valued covariantly constant scalar $\xi^{i}$,%
\[
D_{\mu}\xi^{i}=0\ ,
\]
where $D_{\mu}$\ is the $SU(2)$ covariant derivative. Then, with this
covariantly constant scalar one can construct fluxes which are conserved in
the ordinary sense by contracting the field strength (or its dual) with
$\xi^{i}$. In the present case, it is easy to see that the $Y^{i}$ in Eq.
(\ref{spacc}) are covariantly constant with respect to the gauge field in Eqs.
(\ref{hedgehog1.5}) and (\ref{RN00}). Thus, following \cite{abbottdeser} and
\cite{mann}, the charge $Q=Q(Y^{i})$ is the integral over the 2-sphere at
infinity of the non-Abelian magnetic field $B_{\mu}^{i}$ contracted with
$Y^{i}$,%
\[
Q=\frac{1}{4\pi}\int_{S_{\infty}^{2}}B_{\mu}^{i}Y_{i}n^{\mu}=-\frac{1}{8\pi
}\int_{S_{\infty}^{2}}\Pi_{\theta\phi}d\theta d\phi=-\frac{1}{2}\ ,
\]
where $n^{\mu}$ is the unit normal to $S_{\infty}^{2}$ and we used Eqs.
(\ref{dirac1}) and (\ref{dirac4}). On the other hand, the electric charges
vanish identically.

It is interesting to note that if one would compute the non-abelian magnetic
charges $Q_{M}^{i}$ as surface integrals at infinity without projecting the
magnetic field along the covariantly constant scalar $Y_{i}$ as
\begin{equation}
Q_{M}=Q_{M}^{i}\sigma_{i}=\frac{1}{4\pi}\int_{S_{\infty}^{2}}%
F\ ,\label{charge1}%
\end{equation}
one would get a different result. In the case of the field strength in Eqs.
(\ref{RN01}), (\ref{RN02}) and (\ref{spacc}), the above expression reduces to
\begin{equation}
Q_{M}^{i}\sigma_{i}=\frac{1}{2\pi}\sigma_{i}\int_{S_{\infty}^{2}}Y^{i}%
(\theta,\phi)\Pi_{\theta\phi}d\theta d\phi\ ,\label{charge2}%
\end{equation}
and due to the presence of the functions $Y^{i}(\theta,\phi)$ (whose
expressions are in Eq. (\ref{spacc})) the $Q_{M}^{i}$ would vanish for all the
components of the internal $su(2)$ index $i$.

However, here it is more appropriate the first approach. From the physical
point of view, the idea to project the magnetic field along the $Y_{i}$
corresponds to measure the charge with respect to the radial Pauli matrix
defined in Eq. (\ref{meronmeron2}).

\section{\textit{Jackiw-Rebbi-Hasenfratz-'t Hooft} mechanism}

\label{52}

We have already discussed that the physical origin behind the genuine
non-Abelian nature o f the present solution is the non-trivial realization of
spherical symmetry. Namely, even if the energy-momentum tensor is spherically
symmetric, the field strength in Eq. (\ref{RN01}) is not spherically symmetric
since a spatial rotation does not change $\Pi_{\mu\nu}$ (which is the
pull-back of the volume form of the two-sphere) but it \textit{does change}
the unit radial vector $\overrightarrow{Y}$ in Eq. (\ref{spacc}). Unlike what
happens in Abelian sectors in which the field strength is directly spherically
symmetric (see Eq. (\ref{smollwass1})), the present meron field strength is
spherically symmetric only up to an internal $SU(2)$ rotation which
compensates for the spatial rotation in order to keep Eq. (\ref{RN01}) invariant.

To see this one can look at the field strength of the meron field in Eq.
(\ref{RN01}): such curvature is composed by two factors. The first factor is
the field strength of the Dirac monopole which is invariant under spatial
rotations. The second factor however is the combination $\sigma_{r}%
=Y^{i}\sigma_{i}$ which is not invariant under spatial rotations since the
$Y^{i}$ transform as a vector%
\begin{equation}
Y^{i}\rightarrow R_{j}^{i}Y^{j}\ ,\label{spac}%
\end{equation}
where $R_{j}^{i}$ is the element of $SO(3)$ corresponding to the spatial
rotation. Consequently, $\sigma_{r}$ transforms as follows under a spatial
rotation,%
\[
\sigma_{r}\rightarrow R_{j}^{i}Y^{j}\sigma_{i}\neq\sigma_{r}\ .
\]
Hence, in any equation (such as the $SU(2)$ covariant Klein-Gordon and Dirac
equations) in which the field strength or the corresponding gauge potential
appear explicitly as background fields, the orbital angular momentum
$\overrightarrow{l}$ will not be a symmetry operator. On the other hand, it is
possible to compensate the rotation in Eq. (\ref{spac}) with a corresponding
rotation (generated by $R^{-1}$) of the $SU(2)$ generators in such a way to
keep $\sigma_{r}$ invariant. This means that the symmetry operator in any
equation such as the $SU(2)$ covariant Klein-Gordon equation will be the total
angular momentum $\overrightarrow{J}=\overrightarrow{l}+\overrightarrow
{\sigma}$.

It is precisely the \textit{spherical symmetric up to an internal }%
$SU(2)$\textit{\ rotation} which gives rise to the
\textit{Jackiw-Rebbi-Hasenfratz-'t Hooft} mechanism \cite{JR} \cite{tH}
according to which the excitations of a Bosonic field charged under $SU(2)$
around a background gauge field with the above characteristics behave as
Fermions\footnote{An effect which is very similar to the
\textit{Jackiw-Rebbi-Hasenfratz-'t Hooft }mechanism occurrs for Skyrmions
\cite{skyrme} (for a detailed review see \cite{manton}). Indeed, the
excitations around the Skyrme soliton with winding number equal to one can
behave as Fermions.}. Indeed, following exactly the same arguments in
\cite{JR} \cite{tH}, one can analyze the Klein-Gordon equation for a scalar
field $\Phi$ (which will be assumed to belong to the fundamental
representation) charged under $SU(2)$,%
\begin{equation}
g^{\mu\nu}\left(  \nabla_{\mu}-A_{\mu}\right)  \left(  \nabla_{\nu}-A_{\nu
}\right)  \mathbf{\Phi}=0\label{kg1}%
\end{equation}
where $\nabla_{\mu}$\ are the Levi-Civita corresponding to the metric in Eq.
(\ref{bhsolution}) and $A_{\mu}$ is the meron gauge potential in Eq.
(\ref{hedgehog1.5}). Due to the fact that the metric is static, one can
Fourier-expand the scalar field $\mathbf{\Phi}$ with respect to the time%
\begin{equation}
\mathbf{\Phi}=\exp\left(  iEt\right)  \mathbf{\psi}\left(  r,\theta
,\phi\right)  \ .\label{kg2}%
\end{equation}

The clearest way to disclose the above mechanism is to change coordinates from
the Schwarzschild-like in Eq. (\ref{bhsolution}) to Cartesian-like coordinates
as follows:%
\begin{align*}
\delta_{ij}dx^{i}dx^{j}  & =dx^{2}+dy^{2}+dz^{2}=r^{2}d\Omega^{2}+dr^{2} ,\\
r^{2}d\Omega^{2}  & =\delta_{ij}dx^{i}dx^{j}-dr^{2}\ ,\ \ \ r=\sqrt
{\delta_{ij}x^{i}x^{j}}\ ,\\
dr  & =\frac{1}{r}\left(  \delta_{ij}x^{i}dx^{j}\right)  \ ,\ \ dr^{2}%
=\frac{1}{r^{2}}\left(  x^{i}x^{j}dx^{i}dx^{j}\right)  \ .
\end{align*}
In this coordinates system, the metric in Eq. (\ref{bhsolution}) reads
\begin{align}
ds^{2}  & =g_{\mu\nu}dx^{\mu}dx^{\nu}=-\exp\left(  2a\right)  dt^{2}%
+dx^{i}dx^{j}\left[  \delta_{ij}+\left(  \exp\left(  -2a\right)  -1\right)
\frac{x^{i}x^{j}}{r^{2}}\right]  \ ,\label{cartesianmetric1}\\
g^{tt}  & =-\exp\left(  -2a\right)  \ ,\ \ \ g^{ij}=\left[  
\delta_{ij}+h_{2}(r)\frac{x^{i}x^{j}}{r^{2}}\right]  \ ,\ \ \ g^{0j}%
=0\ ,\label{cartesianmetric2}\\
\exp\left(  2a\right)    & =\left(  1-\frac{\kappa M}{8\pi r}+\frac
{4\kappa\lambda^{2}(\lambda-1)^{2}}{e^{2}r^{2}}-\frac{\Lambda r^{2}}%
{3}\right)  \  ,\ h_{2}=-\frac{r\left(  \exp\left(  -2a\right)
-1\right)  }{1+r^{2}\left(  \exp\left(  -2a\right)  -1\right)  }\ ,\nonumber
\end{align}
while the meron gauge potential reads (see Eq. (\ref{standard4}))
\begin{align}
Y^{0}  & =0,\quad Y^{i}=\widehat{x}^{i}\,,\label{cartesianmeron1}\\
\widehat{x}^{1}  & =\frac{x}{r},\quad\widehat{x}^{2}=\frac{y}{r},\quad
\widehat{x}^{3}=\frac{z}{r}\ ,\label{cartesianmeron2}%
\end{align}%
\begin{align}
A_{\mu}  & =i\lambda P_{\mu}^{k}\sigma_{k}\,,\label{cartesianmeron3}\\
P_{\mu}^{k}  & =\varepsilon_{ijk}\widehat{x}^{i}\partial_{\mu}\widehat{x}%
^{j}=\frac{1}{r^{2}}\varepsilon_{ijk}x^{i}\delta^{j\mu}%
\,.\label{cartesianmeron4}%
\end{align}
If one replaces the ansatz in Eq. (\ref{kg2}) into Eq. (\ref{kg1}) one gets an
effective system of coupled effective Schrodinger equations for the components
of $\mathbf{\psi}$. Explicitly, using the above Cartesian-like coordinates
system, it reads:%
\begin{align}
0  & =\left(  \nabla^{i}\nabla_{i}\right)  \mathbf{\psi}-\exp\left(
-2a\right)  E^{2}\mathbf{\psi}-2\left(  i\lambda\right)  \frac{\sigma^{k}%
}{r^{2}}\varepsilon_{ijk}x^{i}g^{jl}\partial_{l}\mathbf{\psi+V\psi
}\ ,\label{almostfinalkg}\\
\mathbf{V}  & =\left(  i\lambda\right)  ^{2}\left[  \delta_{ij}g^{\mu\nu
}\left(  \partial_{\mu}\widehat{x}^{i}\right)  \left(  \partial_{\nu}%
\widehat{x}^{j}\right)  \right]  \mathbf{1}\ ,\nonumber
\end{align}
where $\mathbf{1}$\ is the identity of the gauge group in the fundamental
representation. Using the explicit expression of the inverse metric in Eq.
(\ref{cartesianmetric2}), Eq. (\ref{kg1}) reads%
\begin{equation}
0=\left(  \nabla^{i}\nabla_{i}\right)  \mathbf{\psi}-\exp\left(  -2a\right)
E^{2}\mathbf{\psi}-2\left(  i\lambda\right)  \frac{1}{r^{2}}\sigma
^{k}\varepsilon_{ijk}x^{i}\partial_{j}\mathbf{\psi+V\psi}\ .\label{finalkg}%
\end{equation}
Indeed, the above equation has exactly the form required for the realization
of the \textit{Jackiw-Rebbi-Hasenfratz-'t Hooft} mechanism since the operator%
\[
\sigma^{k}\varepsilon_{ijk}x^{i}\partial_{j}%
\]
appearing in Eq. (\ref{finalkg}) can be written as%
\[
\sigma^{k}\varepsilon_{ijk}x^{i}\partial_{j}=\overrightarrow{\sigma}%
\cdot\overrightarrow{l}\ ,\ \ \ \overrightarrow{l}=\overrightarrow{r}%
\times\overrightarrow{\nabla}\ ,
\]
($\overrightarrow{\nabla}$ being the flat gradient) which is nothing but the
"isospin-orbit" term able to shift the eigenvalue of the angular momentum from
integer to half-integer. Hence, the final form of Eq. (\ref{finalkg}) is%
\begin{equation}
0=\left(  \nabla^{i}\nabla_{i}\right)  \mathbf{\psi}-\exp\left(  -2a\right)
E^{2}\mathbf{\psi}-2\left(  i\lambda\right)  \frac{1}{r^{2}%
}\overrightarrow{\sigma}\cdot\overrightarrow{l}\mathbf{\psi+V\psi
}\ .\label{finalkg1}%
\end{equation}
In particular, in the asymptotically flat case and in the approximation in
which $r$ is very large Eq. (\ref{finalkg}) reduces to%
\[
\left(  \nabla^{i}\nabla_{i}-a(r)\left(  \overrightarrow{\sigma}%
\cdot\overrightarrow{l}\right)  -b(r)+E^{2}\right)  \mathbf{\psi}%
\mathbf{=}0\ ,
\]
where the explicit forms of $a(r)$ and $b(r)$ are not important as far as the
present discussion is concerned. Hence, the Schrodinger equations are only
invariant under a spatial rotation plus an internal $SU(2)$ rotation generated
by the total angular momentum operator $\overrightarrow{J}$. Thus, the
eigenvalues of the total angular momentum operator $\overrightarrow{J}$ are
good quantum numbers. The key observation in \cite{JR} \cite{tH} is that the
eigenvalues of $\overrightarrow{J}$ can be both integers and half-integers so
that the excitations of $\mathbf{\Phi}$ can behave as Fermions and should be
quantized accordingly. This phenomenon is very interesting in the asymptotic
region when $r$ is very large. In this case the background metric is
approximately flat or (A)dS so that the fields can be quantized using the
standard techniques. In particular, in the asymptotically AdS case which is
relevant in the AdS/CFT correspondence, one can have Fermionic excitations
charged under the gauge group on the boundary without having any Fermionic
field in the bulk.

\section{Conclusions and perspectives}

In the present paper we have constructed a non-Abelian black hole
configuration for the $SU(2)$ Einstein Yang-Mills theory. Even if the metric
coincides with the magnetic Reissner-Nordstrom black hole in which, however,
the coefficient of the $1/r^{2}$ term is not an integration constant, the
solution is intrinsically non-Abelian since it can not be transformed to an
Abelian sector by any globally defined gauge transformation. The gauge field
of the solution has the form of a meron.

An important feature of the present black hole solution when compared with
solutions of Abelian sectors is that the present black hole-meron
configuration is spherically symmetric\textit{\ only up to an internal
}$SU(2)$\textit{\ rotation}. A consequence of this is the realization of the
\textit{Jackiw-Rebbi-Hasenfratz-'t Hooft} mechanism according to which
excitations of Bosonic fields charged under the gauge group can behave as
Fermionic excitations.

The present results can be quite relevant in the context of the AdS/CFT
correspondence since one could have Fermionic excitations at the boundary
without having any Fermionic field in the bulk. It would be interesting to
further explore the consequences of the \textit{Jackiw-Rebbi-Hasenfratz-'t
Hooft} effect within the context of the AdS/CFT correspondence. For instance,
a nice issue to analyze is how to distinguish, by just looking at the boundary
theory, a Fermionic excitation coming from a Fermionic field in the bulk from
a Fermionic excitation generated with the \textit{Jackiw-Rebbi-Hasenfratz-'t
Hooft} effect as it has been proposed here.

It is well known that on flat spaces merons play an important but indirect
role in providing a qualitative explanation of confinement as building blocks
of the instantons. However, they cannot be observed directly due to their
singularities. Indeed, one of the most interesting results of the present
analysis is that the meron singularity is hidden behind the black hole horizon
and, consequently, in a gravitational context, merons could be observed
directly in principle.

The Yang-Mills equations in the present black hole solution force the
proportionality factor of the meron to be $\lambda=1/2$. In the Euclidean
solutions of de Alfaro, Fubini and Furlan \cite{merons1}, the value
$\lambda=1/2$ is related to the fact that merons behave as half-instantons. It
would be interesting to find an analogous interpretation in the Lorentzian case.

\section{Acknowledgments}

{\small {The authors would like to thank Andr\'es Anabal\'on and Gast\'on
Giribet for discussions and useful comments. This work is partially supported
by FONDECYT grants 1120352, 1110167, and 11090281, and by the
\textquotedblleft Southern Theoretical Physics Laboratory\textquotedblright%
\ ACT-91 grant from CONICYT. The Centro de Estudios Cient\'{\i}ficos (CECs) is
funded by the Chilean Government through the Centers of Excellence Base
Financing Program of CONICYT. F. Canfora and F. Correa are also supported by
Proyecto de Insercion CONICYT 79090034 and 79112034. J.O thanks also the
support of Becas Chile Postdoctorales, CONICYT, 2012. }}

\end{document}